\documentclass[runningheads]{llncs}
\usepackage[T1]{fontenc}
\usepackage{graphicx,verbatim}
\usepackage{multirow}
\usepackage{arydshln} 
\usepackage{subfig} 
\usepackage{xcolor} 
\usepackage{graphicx}
\usepackage{amssymb}
\usepackage{algorithm}
\usepackage{algpseudocode}
\usepackage{adjustbox} 
\usepackage{caption}
\usepackage{hyperref}
\usepackage{url}
\usepackage{soul}
\usepackage{amsmath}    
\usepackage{multirow}   
\usepackage{booktabs}   
\usepackage{graphicx}
\usepackage{subcaption}

\begin{document}
\title{\textbf{{M$^{3}$HL}}: Mutual Mask Mix with High-Low Level Feature Consistency for Semi-Supervised Medical Image Segmentation}
\titlerunning{M$^{3}$HL}

\author{Yajun Liu\inst{1} \and Zenghui Zhang \inst{1}\thanks{Corresponding author} \and Jiang Yue \inst{2} \and Weiwei Guo \inst{3} \and Dongying Li \inst{1}}  
\authorrunning{Y. Liu et al.}
\institute{Shanghai Key Laboratory of Intelligent Sensing and Recognition, Shanghai Jiao Tong University, Shanghai, China 
\email{\{liuyajun,zenghui.zhang,dongying.li\}@sjtu.edu.cn} \and
Department of Endocrinology and Metabolism, Renji Hospital, School of Medicine, Shanghai Jiao Tong University, Shanghai, China \\
\email{rjnfm3083@163.com}\and
Center for Digital Innovation, Tongji University, Shanghai, China \\
\email{weiweiguo@tongji.edu.cn}
}
    
\maketitle              
\begin{abstract}

Data augmentation methods inspired by CutMix have demonstrated significant potential in recent semi-supervised medical image segmentation tasks. However, these approaches often apply CutMix operations in a rigid and inflexible manner, while paying insufficient attention to feature-level consistency constraints.
In this paper, we propose a novel method called \textbf{M}utual \textbf{M}ask \textbf{M}ix with \textbf{H}igh-\textbf{L}ow level feature consistency (\textbf{M$^{3}$HL}) to address the aforementioned challenges, which consists of two key components: 
1) M$^{3}$: An enhanced data augmentation operation inspired by the masking strategy from Masked Image Modeling (MIM), which advances conventional CutMix through dynamically adjustable masks to generate spatially complementary image pairs for collaborative training, thereby enabling effective information fusion between labeled and unlabeled images.
2) HL: A hierarchical consistency regularization framework that enforces high-level and low-level feature consistency between unlabeled and mixed images, enabling the model to better capture discriminative feature representations.
Our method achieves state-of-the-art performance on widely adopted medical image segmentation benchmarks including the ACDC and LA datasets. 
Source code is available at \url{https://github.com/PHPJava666/M3HL}.


\end{abstract}

\section{Introduction}
\label{sec:intro}
Semi-supervised medical image segmentation (SSMIS) aims to achieve performance comparable to fully supervised methods while utilizing only limited annotated data, effectively alleviating the challenges of scarce labeled data and labor-intensive annotation processes in medical imaging, which holds significant implications for computer-aided diagnosis and clinical applications. 
Currently, one effective category of SSMIS methods is based on consistency regularization\cite{li2020shape_SASSNet,tarvainen2017mean,wu2022mutual_MCNet+,yu2019uncertainty_UAMT}. These methods enforce consistency constraints to ensure performance stability across different input views, which are typically generated through diverse data augmentation strategies\cite{Bai_2023_CVPR_BCP,he2024pair} or different network initialization approaches\cite{lei2022semi_ASENet,luo2021semi_DTC}.

Among data augmentation based SSMIS methods, BCP\cite{Bai_2023_CVPR_BCP} breaks the paradigm of training labeled and unlabeled data separately, inspired by CutMix\cite{yun2019cutmix} in semi-supervised learning, by generating new training samples through bidirectional copy-pasting of co-located image patches. PSC\cite{he2024pair} extends BCP's approach by splitting paired labeled/unlabeled images into equal-sized patches and randomly shuffling them to create mixed samples. OMF\cite{liu2024overlay} crops foreground and background regions along segmentation edges using label guidance and swaps them across images to synthesize hybrid samples. ABD\cite{chi2024adaptive} enhances segmentation in low-confidence regions via confidence-guided bidirectional replacement of image patches between strongly and weakly augmented inputs. 
These methods, by locally mixing or perturbing labeled and unlabeled data, break the conventional paradigm of independent training, promoting cross-distribution information interaction.
However, these methods rely on rigid and inflexible data mixing strategies (e.g., fixed patch sizes, predefined replacement rules), limiting their adaptability to complex anatomical variations. 
Moreover, they do not incorporate feature-level consistency constraints, which can hinder the effective capture of high-level semantic information, potentially causing error propagation due to local noise, and ultimately restricting the model's ability to capture subtle pathological features.

In this work, inspired by the Masked Image Modeling (MIM)\cite{he2022masked} paradigm in visual representation learning, we propose a dynamic mutual mask mix strategy to refine existing data augmentation frameworks, incorporating high-low level consistency constraints that enable simultaneous attention to global contextual patterns and localized structural details, thereby improving semantic alignment and feature robustness.
Specifically, we first devise a random mask generator that parametrically controls mask patch sizes and mask ratios, enabling dynamic and random mutual mask mixing between labeled and unlabeled data. This dynamic mixing mechanism systematically explores the impact of diverse spatial-contextual combinations on feature learning, compelling the model to develop a more comprehensive understanding of anatomical structures through alternating occlusion and recombination strategies.
Furthermore, we introduce high-low level feature consistency constraints: at the low-level feature space, we enforce geometric consistency of local edge features by constructing multi-view L1 norm constraints between the mixed samples and the unlabeled samples; at the high-level feature space, we design a symmetric cosine similarity metric, constraining the directional consistency of mixed and unlabeled samples' features in the semantic space from multiple dimensions.
This design effectively filters out outlier noise in pseudo-labels through hierarchical feature calibration, enhancing the model's feature discriminability in complex scenarios, such as occlusion and boundary blurring.

In summary, the main contributions of this work are as follows: (1) We introduce a novel dynamic mutual mask mixing (M$^3$) strategy, that enhances semi-supervised medical image segmentation through a random mask generator with adjustable mask patch sizes and ratios, addressing the limitations of rigid data mixing in existing methods. (2) We propose a hierarchical high-low level feature consistency framework (HL), significantly improving the model's ability to capture both global contextual patterns and localized structural details while mitigating pseudo-label noise. (3) We achieve state-of-the-art performance on the ACDC\cite{bernard2018deep} and LA\cite{xiong2021global} datasets, demonstrating the efficacy of our M$^3$HL method in handling scarce labeled data and complex anatomical variations.

\section{Method}
\label{sec:method}
\subsection{Problem Setting and Overall Architecture}
\label{subsec: Overall Architecture}

In our semi-supervised segmentation task, we use a labeled dataset \( \mathcal{D}_l \) with \( N \) labeled samples and an unlabeled dataset \( \mathcal{D}_u \) with \( M \) unlabeled samples, where \( X_l \) and \( Y_l \) represent the labeled image and its corresponding labels, respectively.
Notably, the number of unlabeled samples \( M \) significantly exceeds the number of labeled samples \( N \).

\begin{figure*}[ht]
\centering
\includegraphics[width=1\textwidth]{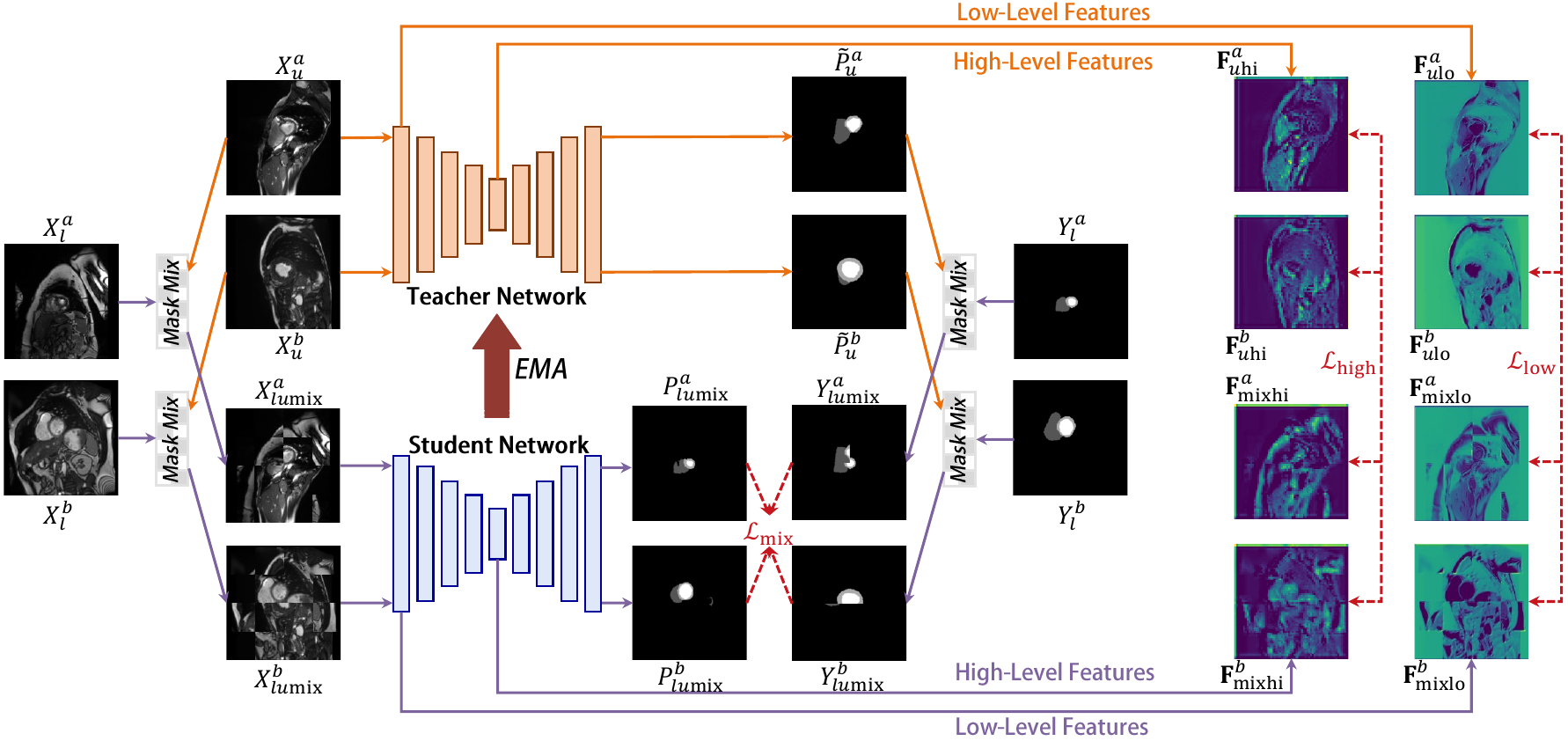}
\caption{Overview of our mutual mask mix with high-low level feature consistency method.}
\label{fig:figure1}
\end{figure*}

The architecture follows a teacher-student paradigm, as shown in Fig.~\ref{fig:figure1}. Each batch of labeled images is divided into two parts, \( X_{l}^a \) and \( X_{l}^b \), and unlabeled images are similarly split into \( X_{u}^a \) and \( X_{u}^b \). After passing \( X_{l}^a \) and \( X_{u}^a \) through our mutual mask mix module, we obtain mixed images \( X_{lu\text{mix}}^a \), and similarly for \( X_{l}^b \) and \( X_{u}^b \) to get \( X_{lu\text{mix}}^b \). These mixed images are input into the student network, while the unlabeled images go into the teacher network. The teacher network's parameters are updated using an exponential moving average (EMA) of the student network. 

The student network generates predictions \( P_{lu\text{mix}}^a \) and \( P_{lu\text{mix}}^b \) for the mixed images, while the teacher network generates pseudo-labels \( \tilde{P}_{u}^a \) and \( \tilde{P}_{u}^b \) for the unlabeled images. These pseudo-labels are mixed with labeled data to generate the mixed outputs \( Y_{lu\text{mix}}^a \) and \( Y_{lu\text{mix}}^b \), which are used in the mutual mask mix loss function $\mathcal{L}_{\text{mix}}(P_{lu\text{mix}}^a, P_{lu\text{mix}}^b, Y_{lu\text{mix}}^a, Y_{lu\text{mix}}^b).$

The teacher network extracts high-level and low-level features from \( X_{u}^a \) and \( X_{u}^b \), denoted as \( \mathbf{F}_{u\text{hi}}^a \), \( \mathbf{F}_{u\text{lo}}^a \), \( \mathbf{F}_{u\text{hi}}^b \), and \( \mathbf{F}_{u\text{lo}}^b \), respectively. Similarly, the student network extracts high-level and low-level features from \( X_{lu\text{mix}}^a \) and \( X_{lu\text{mix}}^b \), denoted as \( \mathbf{F}_{u\text{hi}}^a \), \( \mathbf{F}_{u\text{lo}}^a \) , \( \mathbf{F}_{u\text{hi}}^b \), and \( \mathbf{F}_{u\text{lo}}^b \), respectively. 
The high-level and low-level feature consistency losses are defined as \(\mathcal{L}_{\text{high}}(\mathbf{F}_{u\text{hi}}^a, \mathbf{F}_{u\text{hi}}^b, \mathbf{F}_{\text{mixhi}}^a, \mathbf{F}_{\text{mixhi}}^b)\) and \(\mathcal{L}_{\text{low}}(\mathbf{F}_{u\text{lo}}^a, \mathbf{F}_{u\text{lo}}^b, \mathbf{F}_{\text{mixlo}}^a, \mathbf{F}_{\text{mixlo}}^b)\), respectively, the high-low level feature consistency loss is the sum of both, expressed as $\mathcal{L}_{\text{HL}} = \mathcal{L}_{\text{high}} + \mathcal{L}_{\text{low}}$.

The overall loss function for the training process is composed of the mask mix loss and the weighted high-low level feature consistency loss, expressed as: 

\begin{equation}
    \mathcal{L} = \mathcal{L}_{\text{mix}} + \lambda \mathcal{L}_{\text{HL}} 
    \label{equation:total_loss}
\end{equation}
where the hyperparameter $\lambda $ controls the strength of the high-low level feature consistency constraint. The following section will describe loss functions $\mathcal{L}_{\text{mix}}$ and $\mathcal{L}_{\text{HL}}$ in detail.
It is worth noting that our method does not require a separate supervised loss $\mathcal{L}_{\text{sup}}(X_l, Y_l)$ based on labeled data to train the student network, nor does it require pretraining. We argue that pretraining on very few labeled samples may cause the model to exhibit confirmation bias, which will be verified in our results section.

\subsection{Mutual Mask Mix}
\label{subsec: M3}
\begin{figure*}[ht]
\centering
\includegraphics[width=0.65\textwidth]{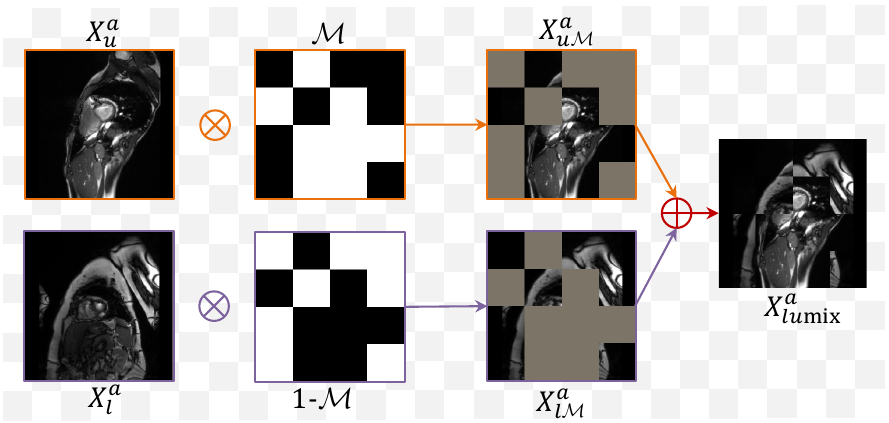}
\caption{Illustration of our mutual mask mix operation.}
\label{fig:figure2}
\end{figure*}

We illustrate the detailed process of the mutual mask mixing operation using the example of generating \( X_{lu\text{mix}}^a \) from \( X_{u}^a \) and \( X_{l}^a \). 
As shown in Fig.~\ref{fig:figure2}, we first generate a mask of the same size as \( X_{u}^a \), for instance, \( 256 \times 256 \). 
The size of the mask patch and the mask ratio are adjustable. In this case, we set the mask patch size to \( 64 \times 64 \) and the mask ratio to 50\%.
Random mask patches are generated within the mask, denoted as \( \mathcal{M} \). By multiplying \( X_u^a \) with \( \mathcal{M} \), we obtain the masked image \( X_{u\mathcal{M}}^a \). Next, we apply the inverse mask \( (1 - \mathcal{M}) \) and multiply it with \( X_l^a \), producing the masked image \( X_{l\mathcal{M}}^a \). Finally, by adding \( X_{u\mathcal{M}}^a \) and \( X_{l\mathcal{M}}^a \), we obtain the mutually mask-mixed image \( X_{lu\text{mix}}^a \).
This process can be mathematically expressed as:    

\begin{equation}
    X_{lu\text{mix}}^a  =  X_{u\mathcal{M}}^a + X_{l\mathcal{M}}^a 
     = X_{u}^a \odot \mathcal{M}  + X_{l}^a \odot (1 - \mathcal{M}) 
\label{equation:X_lumixa}
\end{equation}

We adopt a similar procedure to obtain \( X_{lu\text{mix}}^b \), \( Y_{lu\text{mix}}^a \) and \( Y_{lu\text{mix}}^b \), construct the mutual mask mix loss as follows:

\begin{equation}
\mathcal{L}_{\text{mix}} = \sum_{s \in \{a,b\}} \big(\mathcal{L}_{\text{ce}}(P_{lu\text{mix}}^s, Y_{lu\text{mix}}^s) + \mathcal{L}_{\text{dice}}(P_{lu\text{mix}}^s, Y_{lu\text{mix}}^s)\big) \odot \big(\mathcal{M}  + \alpha(1-\mathcal{M})\big) 
\label{equation:Loss_mix}
\end{equation}

The design of \( \mathcal{L}_{\text{mix}} \), integrating cross-entropy loss \(\mathcal{L}_{\text{ce}} \) and Dice loss \( \mathcal{L}_{\text{dice}} \), and leveraging the dynamic mask \( \mathcal{M} \) and its weighted complement \( (1 - \mathcal{M}) \) (controlled by the parameter \( \alpha \)), optimizes the student model's performance on mixed samples.
This loss function enhances collaborative training between labeled and unlabeled data, strengthens the model's robustness against occlusions, noise, and incomplete data, and improves SSMIS performance through dynamic spatial-contextual exploration, particularly for handling complex anatomical structures and scarce labeled data.

\subsection{High-Low Level Feature Consistency}
\label{subsec: HL}

The low-level features of the mixed samples, \( \mathbf{F}_{\text{mixlo}}^a \) and \( \mathbf{F}_{\text{mixlo}}^b \), obtained after the first downsampling layer in the segmentation network encoder, are constrained by quadruple L1-distance losses with the unlabeled sample's low-level features, \( \mathbf{F}_{u\text{lo}}^a \) and \( \mathbf{F}_{u\text{lo}}^b \), as shown in Eq.\eqref{equation:Loss_low}. Similarly, the high-level features of the mixed samples, \( \mathbf{F}_{\text{mixhi}}^a \) and \( \mathbf{F}_{\text{mixhi}}^b \), extracted after the bottleneck layer of the segmentation network, are constrained with the unlabeled samples' high-level features, \( \mathbf{F}_{u\text{hi}}^a \) and \( \mathbf{F}_{u\text{hi}}^b \)through cosine similarity computation between each pair, as shown in Eq.\eqref{equation:Loss_high}. 
\begin{equation}
    \mathcal{L}_{\text{low}} = \frac{1}{4} \sum_{s \in \{a,b\}} \sum_{t \in \{a,b\}} \| \mathbf{F}_{\text{mixlo}}^s - \mathbf{F}_{u\text{lo}}^t \|_1
    \label{equation:Loss_low}
\end{equation}

\begin{equation}
    \mathcal{L}_{\text{high}} = \frac{1}{4} \sum_{s \in \{a,b\}} \sum_{t \in \{a,b\}} \left[ 1 - \cos\left( \mathbf{F}_{\text{mixhi}}^s, \mathbf{F}_{u\text{hi}}^t \right) \right]
    \label{equation:Loss_high}
\end{equation}

The high-low level feature consistency loss $\mathcal{L}_{\text{HL}}$ is designed to enforce alignment between the mixed and unlabeled samples at both low and high feature levels. Specifically, $\mathcal{L}_{\text{low}}$ minimizes the L1-distance across all pairs of low-level features, ensuring geometric consistency of local edge details. Conversely, $\mathcal{L}_{\text{high}}$ computes the cosine similarity between high-level features to enforce semantic alignment and directional consistency in the semantic space, averaged over all pairs to enhance robustness. Together, $\mathcal{L}_{\text{HL}}$ mitigate noise and improve feature discriminability, particularly in handling complex anatomical structures and incomplete data.

\section{Experiments and Results}
\label{sec:exps}

\subsection{Datasets}
\label{subsec: data}

\textbf{ACDC dataset} - 
The ACDC dataset is a multi-class segmentation dataset that includes the myocardium, left and right ventricles. It consists of 100 cardiac MR imaging samples from 100 patients. We follow the data split in \cite{luo2021semi}, dividing the dataset into training, validation, and test sets with a 70/10/20 ratio.

\noindent\textbf{LA dataset} - 
The LA dataset is a binary segmentation dataset consisting of 100 gadolinium-enhanced MR scans. For consistency, we adopt the data split strategy from \cite{luo2021semi_DTC}, using 80 samples for training and 20  for validation. 
\subsection{Implementation Details and Evaluation Metrics}
\label{subsec: expsetup}
In our experiments, we set the parameters $\lambda $ and $\alpha$ to 0.5. We used an NVIDIA Quadro RTX 6000 GPU (24GB) with a fixed random seed. The SGD optimizer was used with a learning rate of $ 10^{-3} $ and a weight decay of $10^{-4} $. For the LA dataset, we employed a 3D V-Net\cite{milletari2016v} as the backbone network, with input patches randomly cropped to $ 112 \times 112 \times 80 $, a mask patch size of $ 28 \times 28 \times 20 $, a mask ratio of 50\%, a batch size of 8, and a total of 15K training iterations. For the ACDC dataset, we used a 2D U-Net\cite{ronneberger2015u} for segmentation, with input patch sizes of $ 256 \times 256 $, a mask patch size of $ 64 \times 64 $, a mask ratio of 50\%, a batch size of 24, and a total of 30K training iterations. The evaluation metrics included Dice score, Jaccard score, average surface distance (ASD), and 95\% Hausdorff distance (95HD).

\subsection{Comparison with State-of-the-Art Methods}
\label{subsec: Comparison}

\begin{figure*}[ht]
\centering
\includegraphics[width=1\textwidth]{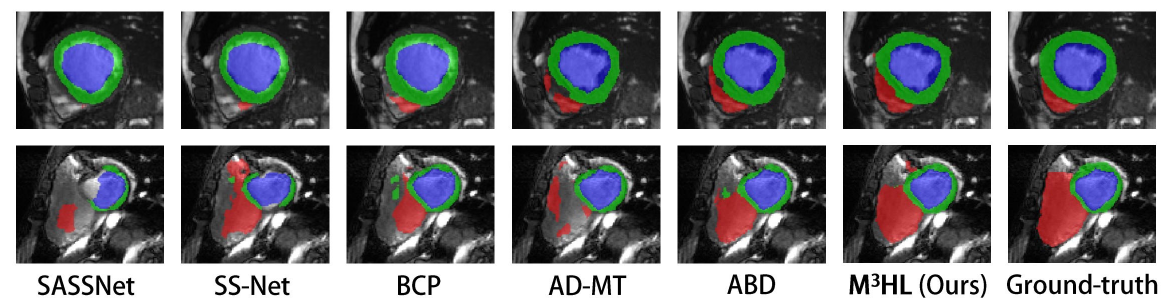}
\caption{Visualization of segmentation results on ACDC dataset with 10\% labeled data.}
\label{fig:ACDC_7label}
\end{figure*}

Table~\ref{tab:merged_results} presents a comprehensive comparison of our proposed M$^{3}$HL method with eight state-of-the-art semi-supervised approaches on the ACDC (10\% labeled) and LA (10\% labeled) datasets. Our method consistently achieves the highest performance across all metrics. Specifically, on the ACDC dataset, M$^{3}$HL outperforms the latest ABD method by 0.66\% in Dice score, 1.28\% in Jaccard, and reduces 95HD and ASD to 1.43 and 0.34, respectively. On the LA dataset, it surpasses the top-performing OMF and AD-MT methods, achieving a Dice score of 91.01\% (0.78\% and 0.46\% improvements over OMF and AD-MT, respectively) and an ASD of 1.59. 
Notably, our method demonstrates robust performance across both datasets without requiring pretraining, unlike BCP and OMF, as this eliminates potential confirmation bias from limited labeled data. 
The qualitative segmentation results in Fig.~\ref{fig:ACDC_7label} show that our method effectively suppresses regions of missegmentation observed in other approaches, producing segmentations closer to the ground-truth.

\begin{table}[ht]
    \centering
    \caption{Segmentation performance comparison on ACDC (10\% labeled) and LA (10\% labeled) datasets. \textbf{*} indicates that the method needs pretraining. \textbf{--} indicates unreported results in original papers.}
    \label{tab:merged_results}
    \resizebox{\textwidth}{!}{%
    \begin{tabular}{c|cccc|cccc}
    \hline
    \multirow{2}{*}{Method} & \multicolumn{4}{c|}{ACDC (10\%/7 labeled)} & \multicolumn{4}{c}{LA (10\%/8 labeled)} \\ 
    \cline{2-9}
    & Dice$\uparrow$ & Jaccard$\uparrow$ & 95HD$\downarrow$ & ASD$\downarrow$ & Dice$\uparrow$ & Jaccard$\uparrow$ & 95HD$\downarrow$ & ASD$\downarrow$ \\ 
    \hline
    U-Net/VNet (SupOnly) &79.41 &68.11 &9.35 &2.70 &82.74 &71.72 &13.35 &3.26 \\ \hline
    UA-MT~\cite{yu2019uncertainty_UAMT}~(MICCAI'19) &81.65 &70.64 &6.88 &2.02 &86.28 &76.11 &18.71 &4.63 \\    
    SASSNet~\cite{li2020shape_SASSNet}~(MICCAI'20) &84.50 &74.34 &5.42 &1.86 &85.22 &75.09 &11.18 &2.89  \\     
    CPS~\cite{chen2021semi}~(CVPR'21) &86.91 &78.11 &5.72 &1.92 &-- &-- &-- &-- \\
    DTC~\cite{luo2021semi_DTC}~(AAAI'21) &-- &-- &-- &-- &87.51 &78.17 &8.23 &2.36 \\
    SS-Net~\cite{wu2022exploring_SSNet}~(MICCAI'22) &86.78 &77.67 &6.07 &1.40 &88.55 &79.62 &7.49 &1.90\\       
    PS-MT~\cite{liu2022perturbed_PSMT}~(CVPR'22) &88.91 &80.79 &4.96 &1.83 &89.72 &81.48 &6.94 &1.92 \\    
    $\text{BCP}^{\textbf{*}}$~\cite{Bai_2023_CVPR_BCP}~(CVPR'23) &88.84 &80.62 &3.98 &1.17 &89.62 &81.31 &6.81 &1.76 \\
    $\text{OMF}^{\textbf{*}}$~\cite{liu2024overlay}(MICCAI'24) &-- &-- &-- &-- &90.23 &82.34 &5.95 &1.63 \\
    AD-MT~\cite{zhao2024alternate}~(ECCV'24) &89.46 &81.47 &1.51 &0.44 &90.55 &82.79 &5.81 &1.70 \\
    ABD~\cite{chi2024adaptive}(CVPR'24) &89.81 &81.95 &1.46 &0.49 &-- &-- &-- &-- \\ \hline
    \textbf{M$^{3}$HL} (Ours) & \textbf{90.47} & \textbf{83.23} & \textbf{1.43} & \textbf{0.34} & \textbf{91.01} & \textbf{83.43} & \textbf{5.72} & \textbf{1.59} \\ 
    \hline
    \end{tabular}%
    }
\end{table}

\subsection{Ablation Studies}
\label{subsec: ablation}

\textbf{Effectiveness of the Proposed Losses $\mathcal{L}_{\text{mix}}$ and $\mathcal{L}_{\text{HL}}$:}
As shown in Table \ref{tab:ablation_study}, we systematically validate the effectiveness of the proposed losses $\mathcal{L}_{\text{mix}}$ and $\mathcal{L}_{\text{HL}}$ by incrementally integrating them with/without the supervised loss $\mathcal{L}_{\text{sup}}$ on the LA dataset (10\% labeled data). The results reveal that either individual or combined use of $\mathcal{L}_{\text{mix}}$ and $\mathcal{L}_{\text{HL}}$ without $\mathcal{L}_{\text{sup}}$ consistently outperforms the baselines with supervised training. Specifically, introducing $\mathcal{L}_{\text{mix}}$ alone achieves significant performance gains (7.02\% and 7.31\% Dice score gains over VNet with/without $\mathcal{L}_{\text{sup}}$, respectively), highlighting the efficacy of our mutual mask mix strategy in fusing semantic information from labeled and unlabeled data through collaborative training. 
Furthermore, incorporating $\mathcal{L}_{\text{HL}}$ on top of $\mathcal{L}_{\text{mix}}$ yields additional improvements, verifying that hierarchical feature utilization enables the model to capture both global contextual and local detailed information for enhanced segmentation.

\begin{table}[ht]
\centering
\caption{Effectiveness of the proposed losses $\mathcal{L}_{\text{mix}}$ and $\mathcal{L}_{\text{HL}}$.}
\label{tab:ablation_study}
\resizebox{0.75\textwidth}{!}{

\begin{tabular}{c|c|c|c|c|c|c|c}
\hline
\multirow{2}{*}{Method} &\multirow{2}{*}{$\mathcal{L}_{\text{sup}}$} & \multirow{2}{*}{$\mathcal{L}_{\text{mix}}$} &\multirow{2}{*}{$\mathcal{L}_{\text{HL}}$} & \multicolumn{4}{c}{Metrics} \\ 
\cline{5-8} 
&&  &  & Dice$\uparrow$ & Jaccard$\uparrow$ & ASD$\downarrow$ & 95HD$\downarrow$\\ 
\hline
VNet&$\checkmark$ & &  & 82.74 & 71.72 & 13.35 & 1.51 \\ 

VNet + M$^{3}$ &$\checkmark$ &$\checkmark$ & & 89.76 & 81.51 & 6.95 & 1.93  \\ 
VNet + HL &$\checkmark$ & & $\checkmark$ & 88.69 & 80.20 & 7.16 & 2.03 \\
VNet + HL + M$^{3}$&$\checkmark$ &$\checkmark$ & $\checkmark$ & 90.32 & 82.48 & 7.06 & 1.68 \\
VNet + M$^{3}$ & &$\checkmark$ & & 90.05 & 82.39 & 7.10 & 1.82   \\ 
VNet + HL & & & $\checkmark$ & 89.40 & 81.43 & 7.28 & 2.02 \\  
\textbf{M$^{3}$HL} (Ours)& &$\checkmark$ & $\checkmark$ & \textbf{91.01} & \textbf{83.43} & \textbf{5.72} & \textbf{1.59} \\ 
\hline
\end{tabular}
}
\end{table}

\begin{figure*}[htb]
\centering
\includegraphics[width=1\textwidth]{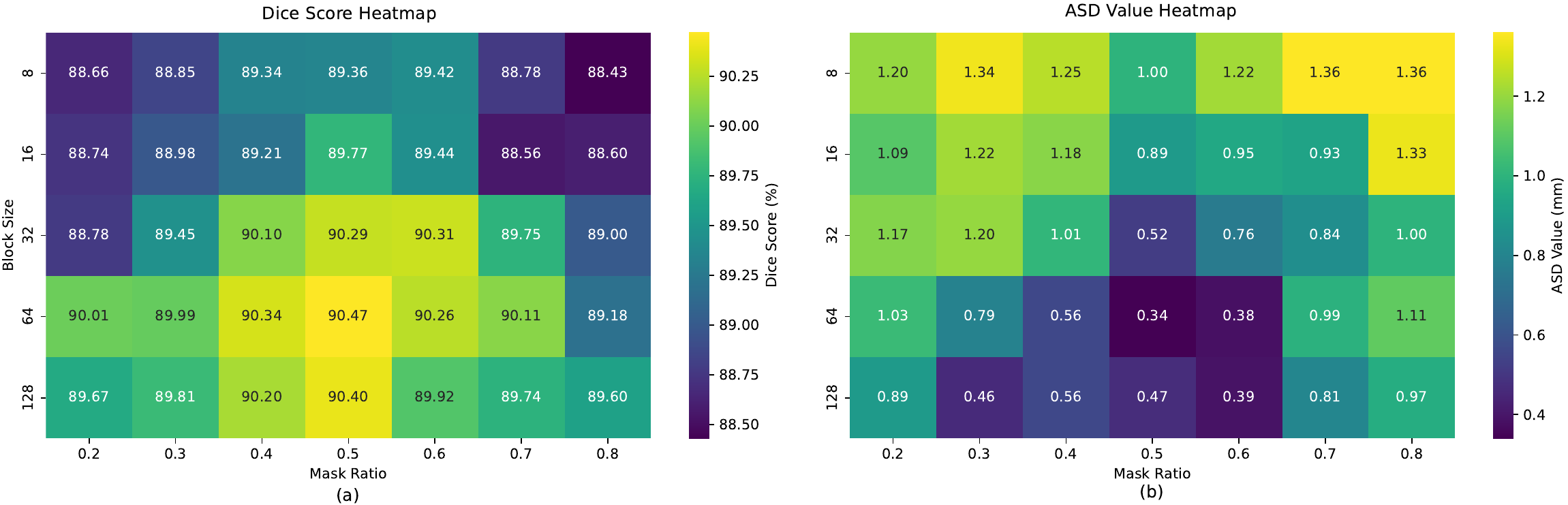}
\caption{Heatmaps of Dice scores and ASD values under varying mask patch sizes and mask ratios.}
\label{fig:dice_asd_heatmaps}
\end{figure*}

\noindent\textbf{Selection of Mask Patch Size and Mask Ratio:}
Fig.~\ref{fig:dice_asd_heatmaps}(a) and Fig.~\ref{fig:dice_asd_heatmaps}(b) present the heatmaps of Dice scores and ASD values under varying mask patch sizes and mask ratios on the ACDC dataset (10\% labeled data). The optimal performance is achieved with a patch size of 64 and a mask ratio of 50\%. This configuration equally masks identical regions in labeled and unlabeled data before mixing, allowing the model to balance feature learning from both data types and achieve optimal representation learning. 

\section{Conclusion}
\label{sec:conclusion}
In this paper, we propose a semi-supervised medical image segmentation method based on mutual mask mix strategy and high-low level feature consistency constraints. The core idea is to enhance data by randomly masking and mutually mixing labeled and unlabeled data, generating mixed data that integrates semantic information from both sources for training. Additionally, by enforcing high-low level feature consistency constraints between mixed samples and unlabeled samples, the method more effectively captures global and local features, thereby improving segmentation performance. In future work, we plan to design more adaptive masking strategies and further explore other feature consistency approaches to address more complex scenarios. 

\section{Disclosure of Interests}
\label{sec:disc_interest}
The authors have no competing interests to declare that are relevant to the content of this article.
\bibliographystyle{splncs04}
\bibliography{mybibliography}
%






\end{document}